# Spontaneous magnetization above $T_C$ in polycrystalline La$_{0.7}$Ca$_{0.3}$MnO$_3$ and La$_{0.7}$Ba$_{0.3}$MnO$_3$

J.A. Turcaud, A.M. Pereira, and K.G. Sandeman
*Department of Physics, Blackett Laboratory, Imperial College London, SW7 2AZ, London, U.K.*

J.S. Amaral
*Departamento de Física and CICECO, Universidade de Aveiro, 3810-193 Aveiro, Portugal*

K. Morrison
*Physics Department, Loughborough University, LE11 3TU, Loughborough, U.K.*

A. Berenov
*Department of Materials, Imperial College London, SW7 2AZ, London, U.K.*

A. Daoud-Aladine
*Science and Technology Facilities Council, Rutherford Appleton Laboratory, Harwell Oxford, OX11 0QX, Didcot, U.K.*

L.F. Cohen
*Department of Physics, Blackett Laboratory, Imperial College London, SW7 2AZ, London, U.K.*

In the present work, spontaneous magnetization is observed in the inverse magnetic susceptibility of La$_{0.7}$Ca$_{0.3}$MnO$_3$ and La$_{0.7}$Ba$_{0.3}$MnO$_3$ compounds above $T_C$ up to a temperature $T^*$. From information gathered from neutron diffraction, dilatometry, and high-field magnetization data, we suggest that $T^*$ is related to the transition temperature of the low-temperature (high magnetic field) magnetic phase. In the temperature region between $T^*$ and $T_C$, the application of a magnetic field drives the system from the high-temperature to low-temperature magnetic phases, the latter possessing a higher magnetization.

## I. INTRODUCTION

Manganites have been extensively studied over many decades, in recent years because of interest in their colossal magnetoresistance (CMR) [1], magnetocaloric [2, 3], and multiferroic [4] properties. To explain the nature of these impressive effects, it was advanced early on that the double-exchange mechanism was the source of magnetic coupling [5, 6] in these materials where localized Mn $t_{2g}$ spins are mediated by itinerant $e_g$ electrons hopping via manganese-oxygen-manganese metallic bonds. To explain, for instance, the large magnetoresistance effect, strong electron-phonon interaction arising from the Jahn-Teller splitting of Mn $d$ levels was added to these early theories to support the drastic change of electronic bandwidth observed at the transition [7]. Further studies concluded that the local distortion revealed by an anomalous thermal expansion between the Curie-Weiss transition temperature $T_C$ and a higher temperature $T^*$, pointed towards the presence of ferromagnetic clusters or polaronic interactions being responsible for the electrical conduction [8-10]. The presence of clusters in the paramagnetic (PM) region was associated with signature magnetic behavior observed in Griffiths phases [11-14].

Kiryukhin *et al.* first showed using neutron scattering that correlated nanoscale lattice distortions are present only in orthorhombic manganite structures, and not in rhombohedral ones [15]. Later Lynn *et al.* demonstrated the existence of a polaron glass phase characterized by short-range polaron correlations, present above $T_C$ [16]. In 2008, de Souza *et al.* [17] suggested that the Griffith-like treatment is inappropriate and a proper description should incorporate the formation of magnetic polarons (manifest as ferromagnetic clusters organized in Mn-spin dimers [18]) coalescing over the temperature range between $T^*$ and $T_C$.

In order to shed new light on the origin of the ordering temperature $T^*$, we have performed a detailed study comparing the magnetic, electronic, and structural properties of La$_{0.7}$Ca$_{0.3}$MnO$_3$ (orthorhombic LCMO) and La$_{0.7}$Ba$_{0.3}$MnO$_3$ (rhombohedral LBMO).

## II. EXPERIMENTAL DETAILS

LBMO and LCMO powders were prepared by solid-state reaction. Stoichiometric amounts of La$_2$O$_3$, MnO$_2$, BaCO$_3$, Y$_2$O$_3$ and CaCO$_3$ were used. Prior to weighing, La$_2$O$_3$ and Y$_2$O$_3$ were calcined for 12 hours at 1000 °C and quenched in a desiccator. MnO$_2$, CaCO$_3$, and BaCO$_3$ were dried at 300 °C for several hours. Starting powders were mixed-ball milled with zirconia balls overnight and calcined in air at 1150 °C for 10 hours. Afterwards, powders were calcined at 1350 °C for 10 hours for up to five times with intermediate grinding (LBMO) or calcined in air at 1150°C for 10 hours, at 1200 °C for 5 hours, at 1250 °C for 5 hours and finally at 1350 °C for 5 hours (LCMO). The resulting powders were rotary milled with zirconia balls at 300 RPM for 3 hours. Powders were uniaxially pressed into 8-mm-wide, 3-mm-thick pellets, at 100 MPa and sintered in air at 1500 °C for 10 hours (LBMO) or 1300 °C for 2 hours (LCMO). Afterwards, samples were oxygenated in air at 900 °C for 50 hours.

Nonsintered powders (LCMO) and ground pellet powders (LBMO) were analyzed by neutron high-resolution powder diffractometer (HRPD) at ISIS, Rutherford Appleton Laboratory. The sample crystal structures were then analyzed by Rietveld refinement. $La_{0.7}Ca_{0.3}MnO_3$ powder crystallized in the orthorhombic perovskite structure (*Pnma*) whereas $La_{0.7}Ba_{0.3}MnO_3$ powder crystallized in the rhombohedral perovskite structure (*R-3CH*).

Magnetometry was measured using a 9 T Quantum Design Physical Property Measurement System (PPMS) fitted with a Vibrating Sample Magnetometer (VSM) option. Thermal expansion and magnetostriction were measured using a capacitance dilatometer [19]. The Seebeck coefficient was continuously measured under a 0.5 K/min cooling rate using a Thermal Transport Option (TTO) mounted on the PPMS.

### III. MAGNETIZATION

Figure 1 shows the field-cooled magnetization response as a function of reduced temperature of both LCMO and LBMO samples under 0.01 and 1 T. The transition temperature $T_C$ is defined as the temperature where the derivative of magnetization with respect to temperature is maximal under a very small field, *i.e.* 0.01 T. The $T_C$ obtained are 258 and 333 K for LCMO and LBMO, respectively.

In the inset of Fig. 1 we see that low-temperature saturation magnetization of LCMO and LBMO are identical, and reach theoretical values of 20.65 A m$^2$/mol, which is equivalent to 3.7 $\mu_B$ per molecular unit formula, *i.e.* per Mn ion. These results show that the samples are magnetically in accordance with what is found in the literature. In contrast the field cooled magnetization in the 1 T field, shows significant differences between samples with the LCMO showing a much sharper change of magnetization associated with the transition.

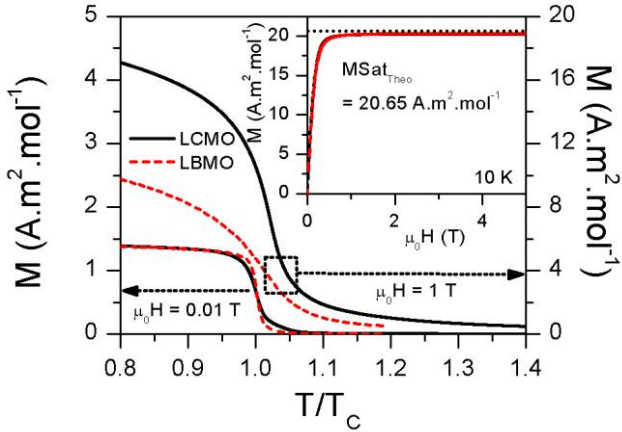

FIG. 1. Main graph: Field-cooled magnetization of LCMO (continuous line) and LBMO (dotted line) as a function of reduced temperature in a 0.01 T (left axis) and a 1 T (right axis) magnetic field. $T_C$ is defined as the temperature where the derivative of magnetization with respect to temperature is maximal under a 0.01 T field. Inset: Saturation magnetization at 10 K for LCMO and LBMO in a magnetic field of 5 T.

Figures 2(a) and 2(b) show the low-field reciprocal susceptibility ($\chi^{-1}$) of LCMO and LBMO, respectively. We observe a deviation from the Curie-Weiss law (kinklike) in the paramagnetic phase. This is the feature that is commonly referred to as the signature for "Griffiths-like" phase behavior [17, 20, 21].

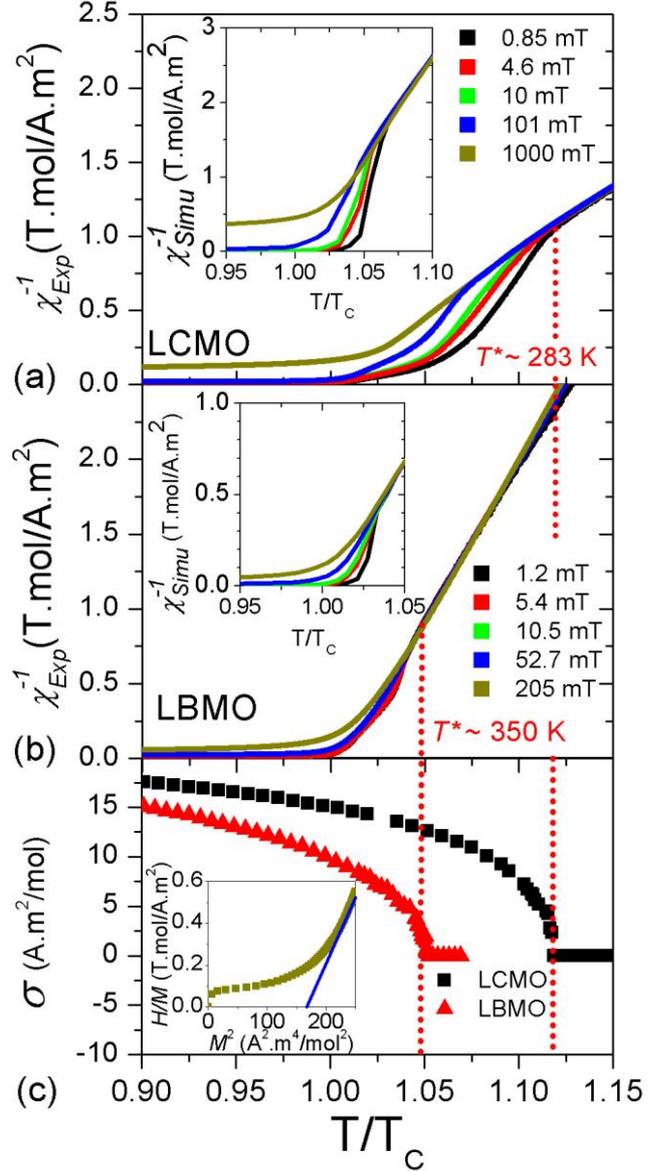

FIG. 2. Main graphs: Experimentally measured field-cooled inverse magnetic susceptibility of LCMO (a) and LBMO (b) as a function of the reduced temperature for increasing applied magnetic field. Insets (a) and (b) are the simulated field-cooled inverse magnetic susceptibility using parameter values for disorder and magnetic clustering from a mean-field and Bean-Rodbell-based analysis via a dedicated software package[i]. (c) Spontaneous magnetization extracted from Arrott plots (inset), as a function of the reduced temperature for both samples. The vertical dashed lines are guides for the eye to indicate for each sample the coincidence of the transition temperature suggested by the spontaneous magnetization extracted from the Arrott plots, and $T^*$ extracted from the inverse susceptibility.

The highest temperature at which this anomalous behavior is present (hereby referred to as $T^*$) is $\sim 1.12 T_C$ for LCMO and ~1.05$T_C$ for LBMO. We also note that the magnetic field required to mask the inverse susceptibility anomalies is about four times higher in the case of LCMO, compared to LBMO.

---
[i] available upon request to jamaral@ua.pt

We can estimate the spontaneous magnetization $\sigma$ by construction of Arrott plots ($H/M$ vs. $M^2$) from M-H loops taken up to 9 T and extrapolation of the intercept with the $M^2$-axis from high-field data [inset to Fig. 2.(c)] [22, 23]. Figure 2(c) shows how the spontaneous magnetization varies with temperature. We note that the temperature where the spontaneous magnetization falls to zero coincides with $T^*$ for both compounds.

## IV. BEAN-RODBELL MODEL AND MAGNETOVOLUME COUPLING

In order to study the nature of the magnetic transition, we have applied the Bean-Rodbell model to our magnetization data. As shown previously [24, 25], systems with second- and first-order phase transitions have been adequately interpreted using this model, which describes in particular the magnetovolume interactions [26]. The model assumes a linear dependence (with a proportionality factor $\beta$) of the Curie temperature ($T_C$) of the system on a relative volume ($v$) change:

$$T_C = T_0 \left[1 + \beta \frac{(v-v_0)}{v_0}\right], \quad \text{(Equation 1)}$$

where $T_0$ is the Curie temperature of the incompressible system.

Considering a material with $K$ compressibility, spin $J$ and $N$ spin density, one defines the $\eta$ parameter:

$$\eta = \frac{5}{2} N k_B K T_0 \beta^2 \frac{[4J(J+1)]^2}{[(2J+1)^4 - 1]}, \quad \text{(Equation 2)}$$

where $k_B$ is the Boltzmann constant. For $\eta > 1$, the transition is considered to be first order, with coupled volume and magnetization discontinuities at specific field and temperature values.

In our work, $K$ and $\beta$ are controlled by the adjustment of $\eta$ manually. The model is a modified form of the Bean-Rodbell model extended to include spin clustering via the parameter $J$. The experimental data can only be well described if a Gaussian distribution of $T_0$ values with variable full width at half maximum (FWHM), accounting for sample inhomogeneity, is incorporated into the model. The parameters $\eta$, $J$, $T_0$, and its FWMH, are tuned in order to provide a best fit to experimental curves such as $M$ vs. $H$, $M$ vs. $T$, and, $H/M$ vs. $M^2$. Figure 3 shows some example data for LBMO and best fit curves. We see a good match especially at high field and high magnetization between measurements and simulated data. As the model assumes a homogeneous and isotropic system, effects such as magnetic domains, anisotropy, and demagnetization are not taken into account, justifying the higher deviation between experimental data and simulations at lower fields. Table I shows, among other results, the parameters obtained from these simulations.

The first-order transition of LCMO is confirmed by its $\eta$ parameter value (>1) [26], in contrast with $\eta$<1 for LBMO. In all other respects the parameters extracted from the model are very similar for the two compounds.

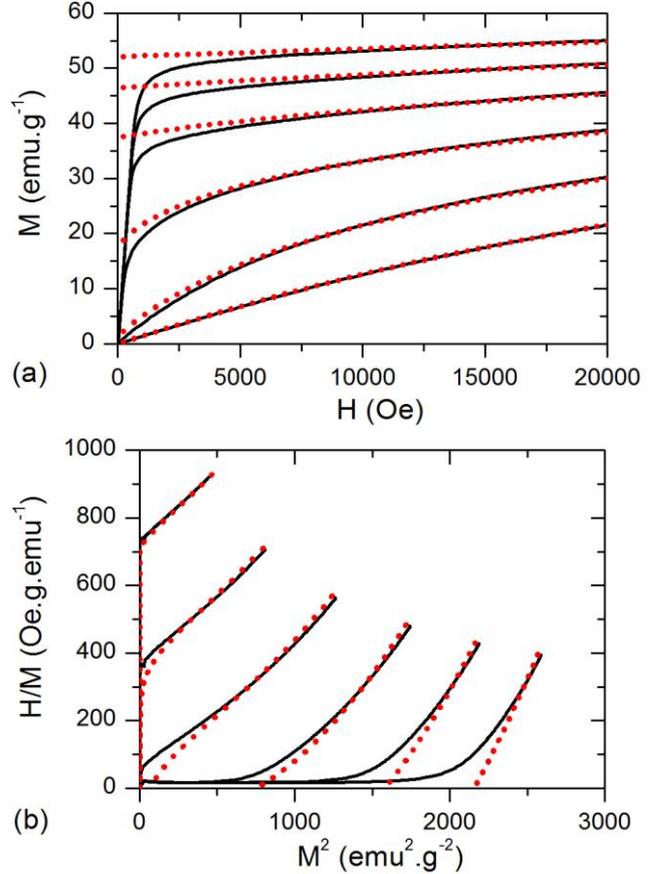

FIG. 3. LBMO experimental (black lines) and simulated (red dots) curves showing (a) magnetization as a function of applied magnetic field, at some representative temperature 300, 310, 320, 330, 340, and 350 K (with lowest temperature at bottom of graph) and (b) typical $H/M$ vs. $M^2$ Arrot plots (for the same temperatures).

The insets of Fig. 2(a) and 2(b) show the numerically simulated isofield temperature dependence of reciprocal susceptibility and demonstrate that the anomalous behavior seen in the reciprocal susceptibility well above $T_C$ is also captured within this simple model. We note that these simulations assume a temperature- and field-independent spin cluster size.

The magnetovolume coupling of both samples was also assessed by measuring the parallel ($\lambda_\parallel$) and perpendicular ($\lambda_\perp$) linear magnetostrictions, relative to an applied magnetic field up to 5 T. For conciseness, only the parallel magnetostriction curves at a few temperatures are shown in Fig. 4(a) for LBMO and in Fig. 4(b) for LCMO. In agreement with the values of $\eta$ found when using the Bean-Rodbell model, we obtained a larger magnetovolume coupling in LCMO than in LBMO. This trend is all the more striking in Fig. 4(c), where the isothermal volume magnetostriction, $\omega$, is plotted for a 1 T field. It is calculated using Eq. (3) [8].

$$\omega = \lambda_\parallel + 2\lambda_\perp \quad \text{(Equation 3)}$$

We also recognize the "S-shape" seen in the magnetostriction curves of LCMO only, which is characteristic of a first-order phase transition and consistent with the values of $\eta$ determined by the above mean-field and Bean-Rodbell-based analysis. These observations confirm that only LCMO possesses a first-order magnetic transition.

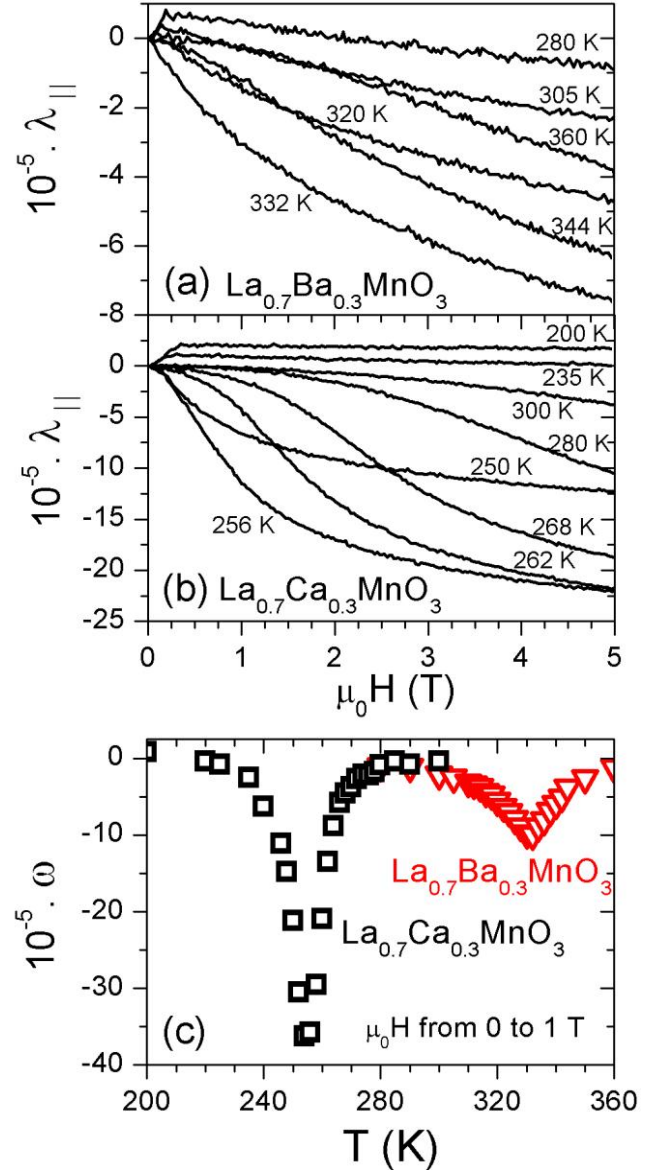

FIG. 4. Linear parallel magnetostriction of LCMO (a) and LBMO (b). (c) Isothermal volume magnetostriction, $\omega$, of LCMO (open black squares) and LBMO (open red triangles) as a function of the temperature under the application of a 1 T magnetic field.

Table I. Theoretical and experimental (10 K) magnetization saturation, parameters extracted from mean-field and Bean-Rodbell-based analysis, from the study of Seebeck coefficient, from polaron refinement, and from neutron diffraction for LCMO and LBMO.

| COMPOSITION | $La_{0.7}Ca_{0.3}$ | $La_{0.7}Ba_{0.3}$ |
|---|---|---|
| **MAGNETIZATION** | | |
| $M_{Sat}$, theo (exp 10 K) (Am²/mol) | 20.65(20.27) | 20.65(20.32) |
| **MEAN-FIELD AND BEAN-RODBELL ANALYSIS** | | |
| $T_C$ (K) | 251.8 | 332.5 |
| $\eta$ | 1.435 | 0.790 |
| Magnetic spin clustering ( no. ions) | 2.89 | 2.58 |
| $T_C$ FWHM/disorder (K) | 11.99 | 11.85 |
| **SEEBECK COEFFICIENT** | | |
| $\Delta E$, activation energy (meV) | - 6.8 | - 8.4 |
| $Q$, electronic cluster size ( no. ions) | 1.4 | 1.3 |
| **NEUTRON DIFFRACTION** | | |
| Space group | *Pnma* | *R-3CH* |
| **0.9 $T_C$** | | |
| a (Å) | 5.47715(6) | 5.53893(6) |
| b (Å) | 5.46002(6) | 5.53893(6) |
| c (Å) | 7.7158(1) | 13.5037(3) |
| Tolerance factor | 0.880(8) | 0.938(12) |
| **1.1 $T_C$** | | |
| a (Å) | 5.4803(1) | 5.54288(6) |
| b (Å) | 5.4644(1) | 5.54288(6) |
| c (Å) | 7.7213(2) | 13.5161(3) |
| Tolerance factor | 0.879(8) | 0.938(12) |

## V. THERMOPOWER

The Seebeck coefficient has also been used to parameterize the electronic behavior in terms of large or small polarons. Figure 5 shows the temperature dependence of the Seebeck coeffient, $S(T)$. At $T_C$, LCMO presents a discontinuous jump of $S(T)$ whereas a smoother change is observed in LBMO. This difference can be interpreted as a jump in the number of heat carriers at the transition itself associated with the observed change of electronic bandwidth [27].

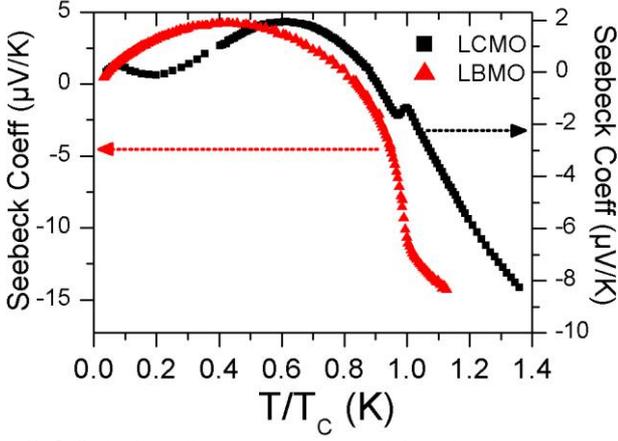

FIG. 5. Seebeck coefficient as a function of the reduced temperature for LBMO (red triangles) and LCMO (black squares). Around $T_C$, a sharp change of the Seebeck coefficient can be seen for LCMO only.

In the paramagnetic regime, $S(T)$, as is well established, does not depend linearly on temperature but shows characteristic polaron-like behavior [27]. We confirmed this by using the polaron model described in detail in Refs. [28-30] which is characterized by the following equation:

$$S = -\frac{1}{|e|}\left(\frac{\Delta E}{T}\right) + S_\infty, \quad \text{(Equation 4)}$$

$$S_\infty = \frac{k_B}{e} \ln\left(\frac{1-Qx}{Qx}\right), \quad \text{(Equation 5)}$$

where $T$, $e$, $S_\infty$, $\Delta E$, $x$, and $Q$ are the temperature, the electron charge, the high temperature limit of the thermopower, the activation energy for hopping, the hole doping concentration, and the size of the polaron, respectively. Results of the fits are shown in Table I.

The negative values obtained for the activation energy in polaron hopping are consistent in amplitude and sign with those reported in Ref. [27]. As explained there, the negative sign can be attributed to hole conduction in the Mn $e_g$ band, as induced by hole doping from replacing trivalent La by divalent Ca or Ba.

Regarding the discrepancy between the cluster and polaron models, the magnetic cluster is determined by a phenomenological model that captures the magnetic correlations between Mn atoms by fitting to the magnetization data above and below $T_C$ and as a simplistic first approximation, the model assumes temperature-independent cluster size and a Gaussian distribution of $T_C$ values. In contrast the electronic cluster size extracted from the thermopower fitting is actually a measure of the size of the polarons which are lattice deformations around each electron. The electronic cluster size was obtained from fitting of the Seebeck coefficient at high temperature, above $T^*$. One would not expect these two different types of fitting to yield the same information, although they reveal similar trends when the two compositions we have studied here are compared.

## VI. ELECTRONIC BANDWIDTH

Radaelli *et al.* [31] showed a direct relationship between the electronic bandwidth and the Curie temperature characterized by Eq. (6), for $A_{0.7}A'_{0.3}MnO_3$ compositions as a function of the average radius between A and A':

$$W \propto \frac{\cos\left[\frac{1}{2}(\pi - \langle Mn|O|Mn\rangle)\right]}{d^{3.5}}. \quad \text{(Equation 6)}$$

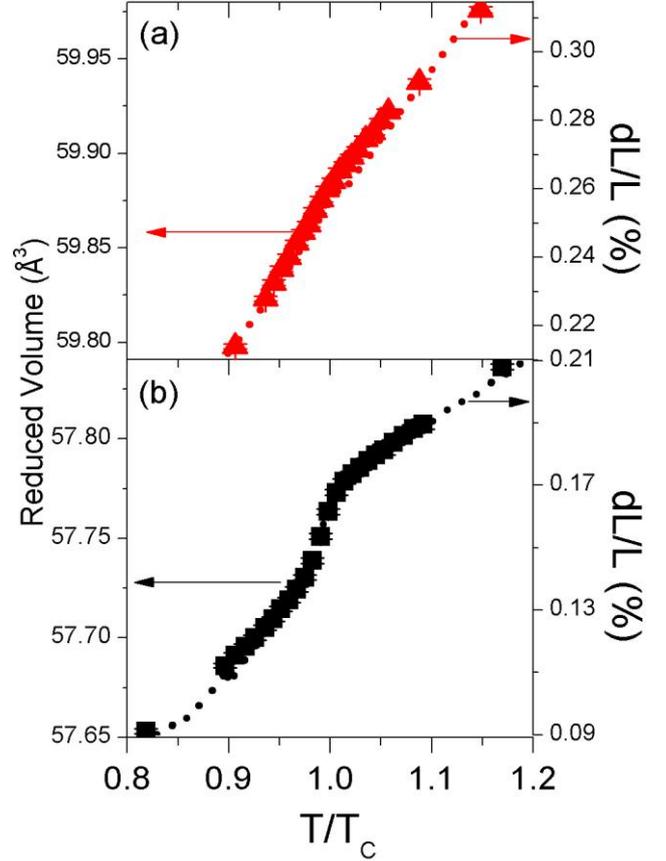

FIG. 6. (a) Refined reduced volume (red triangles and left axis) from neutron diffraction data and thermal expansion (red dotted line and right axis) of LBMO measured with a capacitance dilatometer as a function of reduced temperature. (b) Refined reduced volume (black squares and left axis) from neutron diffraction data and thermal expansion (black dotted line and right axis) of LCMO measured with a capacitance dilatometer as a function of reduced temperature.

Here, $\frac{1}{2}(\pi - \langle Mn|O|Mn\rangle)$ is the "tilt" angle depending on $\langle Mn|O|Mn\rangle$, the Mn-O-Mn bond angle, and $d$ is the Mn-O bond length. Equation (6) implies that if we know both the bond angle and the bond length as a function of temperature, we ought to be able to assign a transition temperature separately to the crystal structure that exists above and below $T_C$. Motivated by this observation, we conducted neutron scattering on our samples to extract detailed structural information as a function of temperature. Data were compared with those extracted from capacitance dilatometry.

Figure 6 shows that the volume changes observed by capacitance dilatometry and calculated from Rietveld refinement neutron diffraction data are in good agreement.

Cell parameters extracted from neutron diffraction data Rietveld refinement are shown in Table I at 0.9 $T_C$ and 1.1 $T_C$.

magnetic transition temperature of the low-temperature phases of LCMO and LBMO.

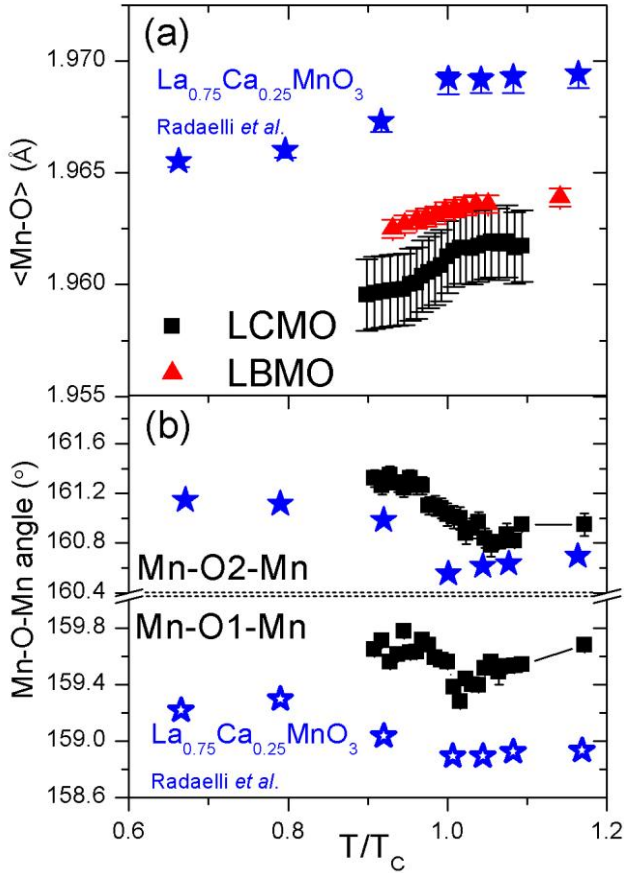

FIG. 7. Average Mn-O bond length (a) and Mn-O-Mn bond angles (b) for LCMO (black squares) and LBMO (red triangles) in comparison with $La_{0.75}Ca_{0.25}MnO_3$ data (blue stars) from the literature [31] as a function of reduced temperature. Mn-O bond length and angles are changing significantly at the transition in LCMO as in $La_{0.75}Ca_{0.25}MnO_3$ and less so in LBMO.

Using Rietveld refinement of our neutron diffraction data, we extracted the average Mn-O bond length and Mn-O-Mn bond angles shown in Fig. 7. The Mn-O bond length changes to a greater extent in LCMO than in LBMO. In addition, if we compare the change in Mn-O bond length seen in a material with similar composition, $La_{0.75}Ca_{0.25}MnO_3$ [31], with the one observed in LCMO, we see that our data show similar trends in amplitude and sign, even though the errors are large. Mn-O-Mn bond angles decrease at the transition in the same manner for LCMO and $La_{0.75}Ca_{0.25}MnO_3$ (Fig. 7). Another feature not shown here is the Jahn-Teller (JT) distortion, calculated using the method described by Radaelli et al. [31]. It is nonexistent in rhombohedral LBMO whereas LCMO and $La_{0.75}Ca_{0.25}MnO_3$ show a similar static JT distortion amplitude of $\sim 5\times 10^{-3}$ Å.

Figure 8 shows a plot motivated by the work of Radaelli et al., displaying the bandwidth evolution determined using Eq. (6) as a function of $T_C$ for $A_{0.7}A'_{0.3}MnO_3$ compositions. Using the values of Mn-O-Mn bond angle and Mn-O bond length from neutron diffraction for temperatures below and above $T_C$, we can use the curve to predict the

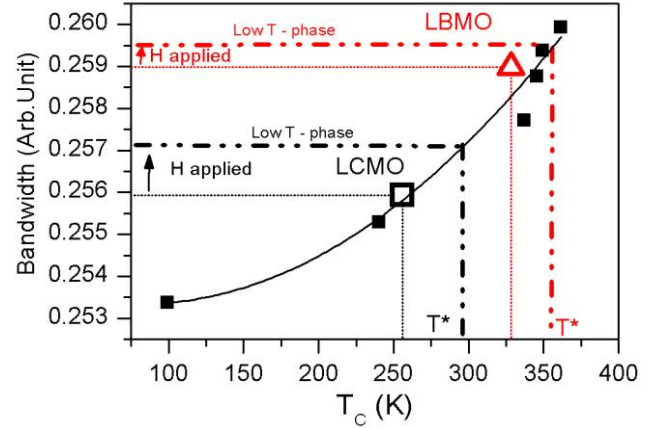

FIG. 8. Graph adapted from Ref. [31] (Radaelli et al.) showing the bandwidths and transition temperatures of $A_{0.7}A'_{0.3}MnO_3$ compounds. The black line is a quadratic fit used to correlate $T_C$ with $W$. LCMO (black square) has been added and LBMO (red triangle) has been identified in this graph using neutron diffraction data above $T_C$ (1.1$T_C$). Bandwidth values for low-temperature phases, calculated using neutron diffraction data below $T_C$ (0.9$T_C$), are shown in dashed lines in black for LCMO and red for LBMO. The transition temperatures of the low-temperature phase of both LCMO and LBMO each coincide with the $T^*$ determined in the Griffiths-like phase using magnetization data at high field and the inverse susceptibility deviation from the Curie-Weiss law in the PM regime.

It is striking to see that transition temperatures of low-temperature phases for both compounds correspond very closely to their respective $T^*$ values that have been extracted previously from spontaneous high-field magnetization studies and inverse susceptibility considerations.

The conclusion from this section supports the earlier spontaneous magnetization data, and suggests that $T^*$ is indeed the $T_C$ of the low-temperature phase.

## VII. DISCUSSION AND CONCLUSIONS

The nature of $T^*$ and its relation with structure, electronic, and magnetic properties are still points of discussion. In fact, Souza et al. [17] referred to $T^*$ as the high temperature limit of the range $T < T_C < T^*$ where ferromagnetic polarons play a dominant role. Polarons are made of local distortions in the lattice, and as such they change the distances between magnetic ions and thus the balance of the exchange magnetic interaction. This interaction is reflected by the appearance of so-called magnetic clustering, and phenomenologically speaking, an apparent spread of $T_C$ values and consequently a deviation in the inverse susceptibility from the Curie-Weiss law.

Here, we have used various models to capture the magnetic and electronic behaviors and to correlate them with the difference in electronic bandwidth that is strongly linked with magnetovolume coupling. We conclude that $T^*$ corresponds to the transition temperature of the low-temperature phase (high-magnetic-field phase). This phase is characterized by a larger bandwidth and is present in the PM phase (above the Curie transition temperature $T_C$) in the form of ferromagnetic clusters.

These new insights contribute to our understanding of the rich and varied magnetic effects observed in manganites.

**ACKNOWLEDGMENTS**

This work was supported by the UK EPSRC Grant No. EP/G060940/1. K.G.S acknowledges financial support from The Royal Society.